\documentclass[twocolumn]{aastex631}
\usepackage{amsmath}

\newcommand{\HTWOO}{H$_{2}\textrm{O}$}

\newcommand{\Mjup}{$M_\mathrm{Jup}$}
\newcommand{\Rjup}{$R_\mathrm{Jup}$}
\newcommand{\Msun}{$M_\odot$}
\newcommand{\Teq}{$T_\mathrm{eq}$}
\newcommand{\kms}{km\,s$^{-1}$}
\newcommand{\K}{KELT-9}
\newcommand{\kp}{$K_\mathrm{p}$}
\newcommand{\vsys}{$v_\mathrm{sys}$}

\newcommand{\caltech}{Department of Astronomy, California Institute of Technology, Pasadena, CA 91125, USA}
\newcommand{\gps}{Division of Geological \& Planetary Sciences, California Institute of Technology, Pasadena, CA 91125, USA}

\newcommand{\ipac}{IPAC, Mail Code 100-22, Caltech, 1200 E. California Boulevard, Pasadena, CA 91125, USA}

\newcommand{\carnegiew}{Earth and Planets Laboratory, Carnegie Institution for Science, Washington, DC, 20015}
\newcommand{\montreal}{Institut Trottier de Recherche sur les Exoplanetes, Universite de Montreal, Montreal, Quebec, H3T 1J4, Canada}
\newcommand{\inaf}{INAF—Osservatorio Astrofisico di Arcetri, Largo Enrico Fermi 5, I-50125 Firenze, Italy}
\newcommand{\oxford}{Department of Physics (Atmospheric, Oceanic and Planetary Physics), University of Oxford, Oxford OX1 3PU, UK}
\newcommand{\hawaii}{Institute for Astronomy, University of Hawai{\'i}, 2680 Woodlawn Drive, Honolulu, HI 96822, USA}
\newcommand{\mitech}{Department of Earth, Atmospheric and Planetary Sciences, Massachusetts Institute of Technology, Cambridge, MA 02139, USA}
\newcommand{\mitkavli}{Kavli Institute for Astrophysics and Space Research, Massachusetts Institute of Technology, Cambridge, MA 02139, USA}
\newcommand{\berkeley}{Department of Astronomy, University of California Berkeley, Berkeley, CA 94720, USA}
\newcommand{\bern}{Weltraumforschung und Planetologie, Physikalisches Institut, University of Bern, Gesellschaftsstrasse 6, 3012 Bern, Switzerland}
\newcommand{\uclagps}{Department of Earth, Planetary, and Space Sciences, University of California, Los Angeles, CA 90095, USA}
\newcommand{\nortredame}{Department of Physics and Astronomy, University of Notre Dame, Notre Dame, IN, 46556, USA}
\newcommand{\ucsb}{Department of Physics, University of California, Santa Barbara, CA 93106, USA}

\shorttitle{}
\shortauthors{Zhang et al.}

\begin{document}

\title{Extreme winds on the emerging dayside of an ultra-hot Jupiter}

\author[0000-0003-0097-4414]{Yapeng Zhang}
\altaffiliation{51 Pegasi b Fellow}
\affiliation{\caltech}

\author[0000-0003-3191-2486]{Joost P. Wardenier}
\affiliation{\bern}
\affiliation{\montreal}

\author[0000-0002-5812-3236]{Aaron Householder}
\altaffiliation{NSF Graduate Research Fellow}
\affiliation{\mitech}
\affiliation{\mitkavli}

\author[0000-0002-9258-5311]{Thaddeus D. Komacek}
\affiliation{\oxford}

\author[0000-0002-3239-5989]{Aurora Y. Kesseli}
\affiliation{\ipac}

\author{Fei Dai}
\affiliation{\hawaii}

\author[0000-0001-8638-0320]{Andrew W. Howard}
\affiliation{\caltech}

\author{Julie Inglis}
\affiliation{\gps}

\author[0000-0002-0531-1073]{Howard Isaacson}
\affiliation{\berkeley}

\author[0000-0002-5375-4725]{Heather A. Knutson}
\affiliation{\gps}

\author{Dimitri Mawet}
\affiliation{\caltech}

\author{Lorenzo Pino}
\affiliation{\inaf}

\author{Nicole Wallack}
\affiliation{\carnegiew}

\author[0000-0002-6618-1137]{Jerry W. Xuan}
\altaffiliation{51 Pegasi b Fellow}
\affiliation{\caltech}
\affiliation{\uclagps}

\author[0000-0001-6416-1274, sname=Carmichael, gname=Theron]{Theron W. Carmichael}
\altaffiliation{NSF Ascend Postdoctoral Fellow}
\affiliation{\hawaii}

\author{Daniel Huber}
\affiliation{\hawaii}

\author{Rena A. Lee}
\affiliation{\hawaii}

\author[0000-0003-2657-3889]{Nicholas Saunders}
\affiliation{Department of Astronomy, Yale University, New Haven, CT 06511, USA}
\affiliation{Institute for Astronomy, University of Hawaii at M\=anoa, 2680 Woodlawn Drive, Honolulu, HI 96822, USA}

\author{Lauren Weiss}
\affiliation{\nortredame}

\author{Jingwen Zhang}
\affiliation{\hawaii}
\affiliation{\ucsb}

\begin{abstract}

High-resolution spectroscopy provides a unique opportunity to directly probe atmospheric dynamics by resolving Doppler shifts of planetary signal as a function of orbital phases. 
Using the optical spectrometer Keck Planet Finder (KPF), we carry out a pilot study on high-resolution phase curve spectra of the ultra-hot Jupiter KELT-9 b. 
We spectrally and temporally resolve its dayside emission from post-transit to pre-eclipse (orbital phase $\phi=0.1-0.45$). The signal strength and width increase with orbital phases as the dayside rotates into view. The net Doppler shift varies progressively from $-13.4\pm0.6$ to $-0.4\pm1.0$ \kms, the extent of which exceeds its rotation velocity of $6.4\pm0.1$ \kms, providing unambiguous evidence of atmospheric winds.
We devise a retrieval framework to fit the full time-series spectra, accounting for the variation of line profiles due to the rotation and winds. We retrieve a supersonic day-to-night wind speed up to $11.7 \pm 0.6$ \kms~on the emerging dayside, representing the most extreme atmospheric winds in hot Jupiters to date.
Comparison to 3D circulation models reveals a weak atmospheric drag, consistent with relatively efficient heat recirculation as also supported by space-based phase curve measurements. 
Additionally, we retrieve the dayside chemistry (including Fe\,{\sc i}, Fe\,{\sc ii}, Ti\,{\sc i}, Ti\,{\sc ii}, Ca\,{\sc i}, Ca\,{\sc ii}, Mg\,{\sc i}, Si\,{\sc i}) and temperature structure, and place constraints on the nightside thermal profile.
Our high-resolution phase curve spectra and the measured supersonic winds provide excellent benchmarks for extreme physics in circulation models, demonstrating the power of this technique in understanding climates of hot Jupiters.

\end{abstract}


\section{Introduction} \label{sec:intro}

Ultra-hot Jupiters (UHJs), which represent the hottest class of close-in gas giant exoplanets with orbital periods less than a few days, are natural laboratories for investigating atmospheric physics under extreme conditions.
UHJs are tidally locked and have extremely irradiated daysides with temperatures above 2200 K, leading to atomic species dominating the spectra \citep{HoeijmakersEtAl2018a, HoeijmakersEtAl2019, ParmentierEtAl2018}. 
Their dayside atmospheres are predicted to be cloud-free, with molecules being thermally dissociated (e.g., H$_2$O/OH) and atoms being further ionized  \citep{LothringerEtAl2018, KitzmannEtAl2018}. 
As atomic metals, metal hydrides, and oxides are efficient absorbers at UV and optical wavelengths, significant heat is deposited in the upper atmospheres, resulting in inverted temperature structure as a function of pressure. The thermal inversion has been ubiquitously detected on the dayside of UHJs, showing spectral features in emission \citep[e.g.,][]{KreidbergEtAl2018, PinoEtAl2020, NugrohoEtAl2020, YanEtAl2020}. 
On the other hand, refractory species may condense out on the nightside and be incorporated into clouds \citep{SpiegelEtAl2009}. The extreme temperature and chemistry gradients across hemispheres make them interesting targets to investigate the three-dimensional nature and atmospheric circulation of planets.

Previous phase curve observations with space telescopes such as Spitzer, TESS, and CHEOPS have enabled mapping the temperature distribution in hot Jupiter atmospheres \citep[e.g.,][]{KnutsonEtAl2007, CowanAgol2008}.
The interplay between various factors, including irradiation, molecular dissociation, disequilibrium chemistry, and magnetic effects, regulates the heat redistribution \citep{BellCowan2018, BeltzEtAl2022, TanKomacek2019}. 
Generally, the UHJ population shows enhanced day-to-night temperature differences and hence smaller heat recirculation efficiencies compared to normal hot Jupiters \citep{BellEtAl2021, DangEtAl2024}. This has been explained by theoretical studies and general circulation models (GCMs), suggesting that the radiative heating/cooling and the frictional drag (due to turbulence or ionized atmospheres coupled to magnetic fields) can efficiently damp waves, leading to inefficient day-to-night heat transport at high equilibrium temperatures \citep{KomacekShowman2016, KomacekEtAl2017, BeltzEtAl2022, BeltzRauscher2024}.
The different circulation patterns also display distinct winds and dynamical signatures. GCMs suggest a regime transition from zonal jets to global day-to-night flows with increasing irradiation and lower pressures \citep{ShowmanEtAl2013, KomacekShowman2016}.
High-resolution spectroscopy can resolve the spectral features which provides a unique opportunity to directly measure atmospheric winds \citep{Snellen2025}. 
Therefore, characterization of dynamics with high-resolution spectroscopy forms a complementary and powerful approach to reveal atmospheric circulations in addition to space-based phase curves.

With the increased signal-to-noise ratio (S/N) enabled by the ensemble of high-resolution spectrographs ($\lambda/\Delta\lambda\sim100,000$) on large telescopes, temporally and spectrally resolving atmospheric features has recently become possible. 
Time-resolved transmission spectroscopy reveals asymmetric absorption signals between the dawn and dusk limbs of UHJs, such as WASP-76 b and WASP-121 b, which show absorption that is progressively blueshifted and gaining in amplitude during the transit \citep{EhrenreichEtAl2020, KesseliSnellen2021, WardenierEtAl2024, SeidelEtAl2025, PrinothEtAl2025}.
The temporal variations can be interpreted as a consequence of day-to-night winds combined with spatial gradients in thermal structure, chemistry, or clouds between day and night sides. 
These observations have been extensively investigated with state-of-the-art GCMs and have led to key insights into the 3D structure of UHJs \citep{BeltzEtAl2022, KesseliEtAl2024, SavelEtAl2022, WardenierEtAl2021, WardenierEtAl2024}.

Although these transmission observations have been vital so far, the region of the atmosphere probed during transmission spectroscopy is limited to the terminators.
Resolving planetary emission as a function of orbital phase angles provides critical information on thermal and dynamical structures across a wide range of longitudes.
Previous high-resolution emission spectroscopy studies explore the dayside orbital phases around secondary eclipses. 
\cite{HermanEtAl2022, vanSluijsEtAl2023} reported asymmetric signal strength between the pre- and post-eclipse phases in WASP-33 b, likely as a result of an eastward hotspot offset.
In contrast, \cite{PinoEtAl2022, Ridden-HarperEtAl2023} found no evidence for variation of the amplitude of spectral lines with phase in the day-side of \K~b.
\cite{PinoEtAl2020} also investigates the Doppler shifts as a function of orbital phases to constrain the wind pattern. They detect a deviation between the Doppler shift of iron lines and the best-fit circular orbit solution of the planet, which they attribute to winds of several up to 10 \kms. They also study the phase dependence of the Doppler shift and favor day-to-night winds over an equatorial jet, but their conclusions on the geometry of winds remain tentative due to the limited S/N.

On the modeling side, there are multiple studies that investigate phase-resolved emission spectra using GCMs, including \cite{ZhangEtAl2017a, BeltzRauscher2024, WardenierEtAl2025}. These predict quasi-sinusoidal Doppler shifts as a function of orbital phases, driven by planet rotation. 
Different atmospheric drags or magnetic effects, showing various wind patterns, can lead to deviations from such curves \citep{WardenierEtAl2025}. In particular, the most prominent Doppler shift occurs in phases before quadrature (orbital phase: $\phi\sim0.05-0.25$), which is therefore the most diagnostic. Yet, no existing observations have high enough S/N and the phase coverage to conclusively constrain different GCM predictions.

In this paper, we perform a case study on one of the most favorable emission targets, \K~b, to resolve the planetary signal across a wide range of orbital phases and unveil its thermal structure and wind pattern. 
We present our observations and data reduction in Section~\ref{sec:observation}. We perform cross-correlation and retrieval analysis to detect the planetary emission and constrain atmospheric properties. The model setup and analysis methods are described in Section~\ref{sec:analysis}. Then, Section~\ref{sec:result} presents the detection and retrieval results. We discuss the comparison to GCMs and implications on the circulation and climate of \K~b in Section~\ref{sec:discussion}. Finally, we summarize the findings in Section~\ref{sec:conclusion}.

\section{Observations and Data Reduction} \label{sec:observation}

\begin{figure*}[t]
\centering
    \includegraphics[width=\linewidth]{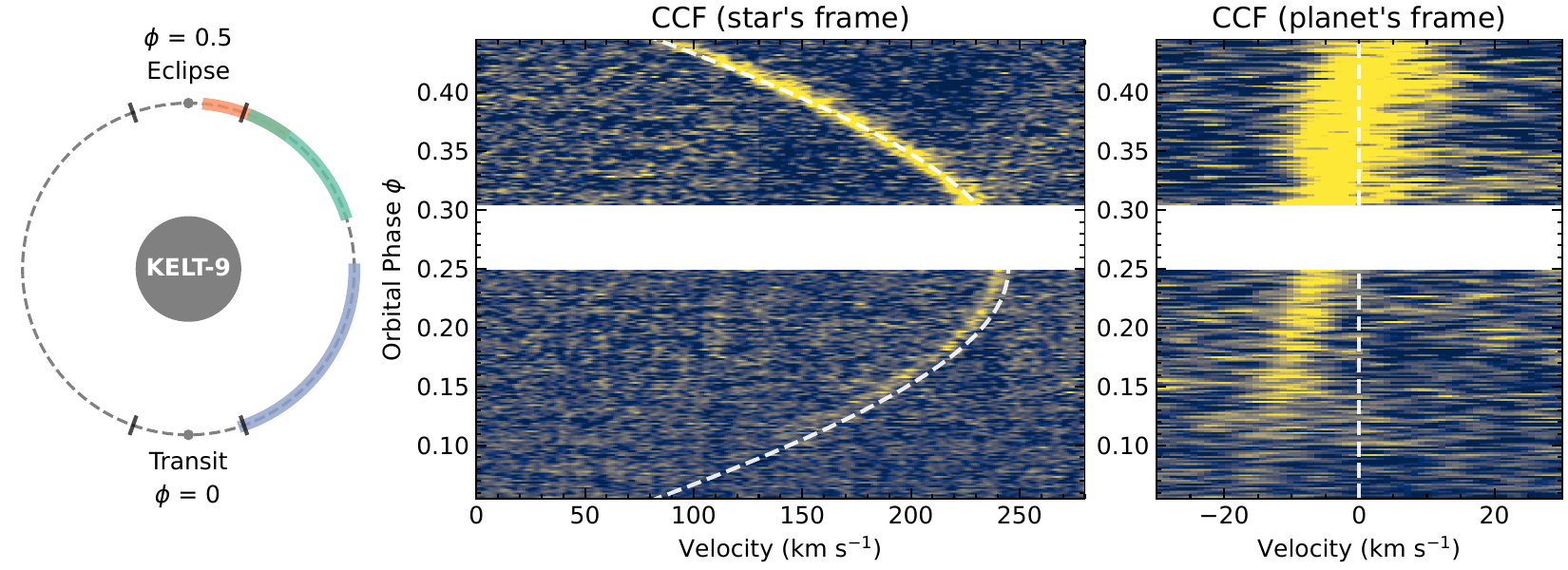}
    \caption{Detection of \K~b's emission as a function of orbital phases. Left panel: phase coverage of our KPF observations in three shaded curves. The black bars mark the timing of the transit and secondary eclipse. Middle panel: CCF map of the spectral data from post-transit to pre-eclipse in star's rest frame. The planetary signal shows as the bright yellow trace following a sinusoidal function  of phases (white dashed line) due to the orbital motion. Right panel: CCF map shifted to the planet's rest frame assuming a orbital semi-amplitude of \kp=244.8 \kms~(i.e., the inferred \kp~value from retrievals as detailed in Section~\ref{sec:result_dynamics}). The planetary emission profile shows clear variation with phases.
    \label{fig:ccf_phase}}
\end{figure*}

\K~b is the hottest exoplanet (\Teq $\sim4000$ K) discovered to date \citep{GaudiEtAl2017}. It has a 1.48-day, nearly polar orbit around an A0 star \citep{GaudiEtAl2017, AhlersEtAl2020}. 
We summarize literature studies on \K~b in Appendix~\ref{app:system}. 
Here we observed emission from \K~b at multiple orbital phases with the Keck Planet Finder \citep[KPF;][]{GibsonEtAl2016, GibsonEtAl2018, GibsonEtAl2020a, GibsonEtAl2024} on UT 2024 August 8, 20, and 27 (PI: Zhang). KPF is an echelle spectrometer that covers the optical wavelength range of 440-870 nm with a spectral resolving power of $\mathcal{R}\sim98\,000$. 
As the planet's orbital semi-amplitude is high ($>230$ \kms), we choose an exposure time of 150 seconds to limit the smearing of the signal within each exposure.
During observations, the airmass varies from 1.07 to 1.7, and the average seeing condition ranges from 0.3\arcsec~on August 8 to 0.5\arcsec~on the August 20 and 27. 
This delivers a typical S/N per pixel of $\sim$400, 330, and 320 at 650 nm for each night, respectively.  
The total on-target time is 3.5h (84 exposures), 2.5h (59 exposures), and 5.2h (125 exposures) each night, respectively. 
The observations cover a wide range of orbital phases between $\phi=0.05-0.25$ and $\phi=0.30-0.48$ as illustrated in Fig.~\ref{fig:ccf_phase}.

We take the spectra extracted with the KPF Data Reduction pipeline \citep[DRP,][]{GibsonEtAl2020a}, which performs flat-fielding, order tracing, and optimal extraction.
The DRP achieves wavelength calibration with laser frequency comb and thorium-argon exposures bracketing the science observations.
We further perform blaze correction using the blaze functions extracted from the smooth lamp pattern (see \cite{HouseholderEtAl2025} for details). Then we correct for telluric absorption features caused mainly by H$_2$O and O$_2$ in the Earth's atmosphere using the ESO sky tool \texttt{molecfit} \citep{SmetteEtAl2015}. 
The details of telluric correction can be found in Appendix~\ref{app:molecfit}.
After telluric correction, we shift the observations to the stellar rest frame by correcting for the systemic, barycentric velocities, and star's reflex motion. 
The systemic velocity is measured from the cross-correlation of the combined stellar spectrum with a PHOENIX model \citep{HusserEtAl2013} of $T_\mathrm{eff} = 10000$ K. 
We fit a Gaussian profile to the cross-correlation function (CCF) and find the center with \vsys$=-17.65 \pm 0.06$ \kms.
We note that it is not the optimal approach for estimating the \vsys~of the fast rotating star. To account for the \vsys~uncertainty, we add a free parameter $\Delta$\vsys~in our following retrieval analysis as detailed in Section~\ref{sec:analysis_retrieval} and \ref{sec:result_dynamics}.

We normalize each spectrum with a third-order polynomial per spectral order and obtained a master stellar spectrum by average combining the time series. Then, individual exposures in the time series are divided by the master stellar spectrum to remove the stellar features. 
Any remaining low-order trend in the continuum is further removed with a fifth-order polynomial.
The noisy wavelength channels due to strong telluric contamination are masked by clipping outliers above 5$\sigma$. 
This leads to $\sim1.2\%$ of wavelength channels being masked.
The observation uncertainties $\sigma_0$ are estimated from the standard deviation of the residuals per wavelength channel after removing the master stellar spectrum.
Given the exquisite data quality, we avoid any additional detrending processing to ensure the planetary signal unaltered.


\section{Method}\label{sec:analysis}

\subsection{Atmospheric Model} \label{sec:model}

To perform cross-correlation and retrieval analysis, we build atmospheric models of \K~b 
and generate synthetic spectra using the radiative transfer tool \texttt{petitRADTRANS} \citep[\texttt{pRT};][]{MolliereEtAl2019}.
We use the analytical pressure-temperature (P-T) profile in radiative equilibrium from \cite{Guillot2010}, 
which involves three parameters: the atmospheric opacity in the infrared (IR) wavelengths $\kappa_\mathrm{IR}$, 
the ratio between the optical and IR opacity $\gamma$, the equilibrium temperature $T_\mathrm{eq}$.
Given the temperature structure, equilibrium chemical abundances are calculated on-the-fly with \texttt{easychem} \citep{LeiMolliere2025}. 
Chemical equilibrium is expected in the lower atmosphere probed by emission spectroscopy,
where the thermal rates dominate over stellar irradiation and advection.
The calculation also takes into account the thermal ionization of neutral atoms such as Mg, Ca, Si, Fe, Ti, and V.
As chemical abundances are not the focus of this study, we fix the composition to solar metallicity [Fe/H] = 0, which is typically retrieved for refractories in UHJs \citep{GandhiEtAl2023a, PelletierEtAl2024}. 
We note that the model did not account for the non-local thermodynamic equilibrium (NLTE) effects on the thermal structure and line formation \citep{TurnerEtAl2020, FossatiEtAl2021}. 
However, the difference is expected to be small for metal emission lines originated from the lower atmosphere ($P>10^{-4}$ bar).

We generate emission spectra using \texttt{pRT} with the line-by-line mode. 
We consider the continuum opacity sources of collision-induced absorption from H$_2$-H$_2$, H$_2$-He, and H$^-$ bound-free and free-free absorption, 
as well as line opacities from
Fe\,{\sc i}, Fe\,{\sc ii}, Ti\,{\sc i}, Ti\,{\sc ii}, Na\,{\sc i}, Ca\,{\sc i}, Ca\,{\sc ii}, Mg\,{\sc i}, Si\,{\sc i}, V\,{\sc i}, and V\,{\sc ii}.
The line opacity cross-sections are computed from \texttt{Kurucz} line lists \citep{Kurucz2018} using the Python package \texttt{pyROX} \footnote{\url{https://github.com/samderegt/pyROX}} \citep{deRegtEtAl2025a}
for temperatures up to 9000 K, which is sufficient for the temperature range of \K~b.
To speed up the calculation, we downsample the original opacity tables of $\mathcal{R} = 10^6$ by a factor of 3. 
This downsampling factor has been benchmarked in high-resolution retrieval studies \citep{XuanEtAl2022, ZhangEtAl2024}. 
Then we apply instrument broadening by convolving the model spectrum to the KPF spectral resolution of $\mathcal{R}=98,000$ with a Gaussian kernel. 
The planet-to-star flux ratio is then computed by $(R_\mathrm{p}/R_\ast)^2F_\mathrm{p}/F_\ast$, 
where we use a PHOENIX model of $T_\mathrm{eff} = 10000$ K broadened by $v\sin i=115$ \kms~as the stellar spectrum.
Finally, we remove the continuum from the model by subtracting the Gaussian-smoothed version with a kernel size of 150 pixels ($\sim2-4$ \AA).

\subsection{Cross-correlation Analysis} \label{sec:analysis_ccf}

We perform cross-correlation analysis to detect the planetary signal in the time series spectra.
The CCF is computed by
\begin{equation}
    \mathrm{CCF}(v) = \mathbf{f}(v)^T \Sigma_0^{-1} \mathbf{y},
\end{equation}
where $\mathbf{f}(v)$ is the model template shifted to a given velocity $v$, 
$\Sigma_0$ is the covariance matrix with the diagonal items populated by
observation uncertainties $\sigma_0$, and $\mathbf{y}$ is the individual spectrum of the time-series observations.
We calculate the CCF for a velocity range of $[-300, 300]$ \kms~with a step of 1 \kms.
We also compute the \vsys-\kp~diagram as commonly presented in the literature \citep{BrogiEtAl2012}.
This is done by shifting the CCF time series to planet's rest frame following $v_\mathrm{p} = K_\mathrm{p} \sin (2\pi\phi)$ and coadding them
assuming a range of \kp~values from 0 to 350 \kms.

In the CCF map, we note a significant residual pattern caused by stellar pulsation (see Fig.~\ref{fig:ccf_pulsation} in Appendix),
as also identified in previous light curve and spectroscopic observations \citep{WongEtAl2020, WyttenbachEtAl2020}.
The pulsation signal prominently contaminates the data within the velocity range of the stellar rotation $v\sin i\sim 115$ \kms. 
At most orbital phases, the planet's RV is larger than this value so that the planetary signal is unaffected. 
However, the stellar contamination overlaps the planetary signal at $\phi<0.1$ and $\phi>0.4$, 
making the measured line strength and shape less reliable at those phases.
For visual clarity of the CCF map shown in Fig.~\ref{fig:ccf_phase}, we use the \texttt{tinygp}\footnote{\url{https://github.com/dfm/tinygp}} package \citep{Foreman-MackeyEtAl2024} to perform  Gaussian Processes (GP) fitting to the CCFs within the velocity range of $\pm115$ \kms~to remove the pulsation signal \citep{WyttenbachEtAl2020}. 
We note that the contaminated phases are excluded for subsequent analyses such as retrievals and phase-combined CCF detections.

\begin{figure*}[t]
\centering
    \includegraphics[width=\linewidth]{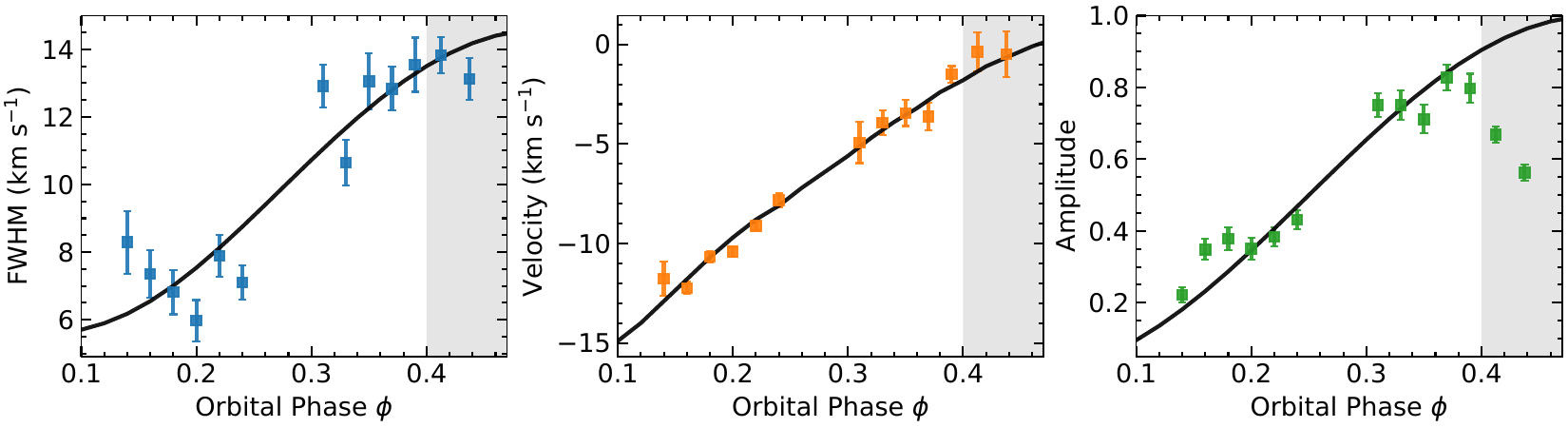}
    \caption{Phase dependency of the line width, net Doppler shift, and amplitude of the \K~b's emission signal as extracted from the CCFs shown in Fig.~\ref{fig:ccf_phase}. The black lines represent the simple harmonic forms adopted in the retrieval analysis (see Eq.\ref{eq:line_width} for the FWHM, Eq.~\ref{eq:line_strength} for the amplitude, and Eq.~\ref{eq:line_doppler} for the net Doppler shift).
    The signal amplitude suffers from stellar contamination for $\phi>0.4$ (gray shaded region).
    \label{fig:phase_variation}}
\end{figure*}

\subsection{Phase-resolved Retrieval Analysis} \label{sec:analysis_retrieval}

To retrieve atmospheric properties of \K~b, we carry out Bayesian inference analysis
on the time-series observations. The large phase coverage presents challenges in fitting all spectra simultaneously, 
as the planetary emission varies significantly across orbital phases. On the other hand, 
it provides an excellent opportunity to reveal the 3D structure of the atmosphere.
Inspired by previous emission studies and GCM results, we parameterize the phase dependency of the spectral features, 
including line strength, width, and Doppler shift, using harmonic forms. 
Following \cite{HermanEtAl2022, Ridden-HarperEtAl2023, HoeijmakersEtAl2024}, the line strength as a function of orbital phases assuming a uniform dayside and nightside is modeled as
\begin{equation} \label{eq:line_strength}
A_{\mathrm{p}}(\phi)= 1-C \cos ^2(\pi(\phi-\theta)) ,
\end{equation}
where $C$ is the contrast factor between the dayside and nightside emissions in the form of $1-F_\mathrm{night}/F_\mathrm{day}$, 
and $\theta$ is the phase offset of the peak emission.
We fix $\theta=0$ because our observations do not cover both pre- and post-eclipse phases, and therefore are less sensitive to this parameter.  
We fix $C=1$ for considering the dayside contribution alone, i.e. $A_\mathrm{day} = 1- \cos ^2(\pi\phi)$. 
The nightside contribution is taken into account separately with a $0.5-\phi$ phase shift in Eq.~\ref{eq:line_strength}, which becomes $A_\mathrm{night} = 1- \sin ^2(\pi\phi)$.
Although these are simplified prescriptions, the observed phase dependence of the amplitude matches well with this simple form as shown in Fig.~\ref{fig:phase_variation}.
Here we impose the phase-dependent scaling of the signal without any free parameters. In practice, any uncertainties related to the model amplitude will be effectively incorporated into the retrieved P-T profile.

Similarly, the phase variation of line widths is modeled as
\begin{equation} \label{eq:line_width}
    \mathrm{FWHM}(\phi)=\omega\sin ^2(\pi\phi),
\end{equation} 
where $\omega$ is the maximum line width when the dayside hemisphere is fully in view. 
This is informed by the GCM prediction \citep{WardenierEtAl2025}, and the measured line widths from our observations also support this as shown in Fig.~\ref{fig:phase_variation}. 

As for the net Doppler shift, its phase variation is mainly driven by the planet's rotation ($v_\mathrm{eq}$) and additional atmospheric winds.
To describe winds flowing from the substellar point towards the nightside, we assume a simple sinusoidal dependence of longitudes as 
$v_\mathrm{wind}(\varphi) = u_\varphi \sin\varphi$, where $u_\varphi$ is the maximum zonal wind speed near terminators. 
It means that the wind is weak near the substellar point and reaches its maximum near the terminators.
This prescription is motivated by the day-to-night flow pattern in GCM simulations (see Section~\ref{sec:gcm} and Fig.~\ref{fig:u_phi}).
The line-of-sight velocity for each longitude ($\varphi$) and latitude ($\theta$) location on the planet is given by
\begin{equation} \label{eq:line_doppler}
    v_\mathrm{los}(\varphi, \theta) = (v_\mathrm{eq} + u_\varphi\sin\varphi) \sin(\phi+\varphi) \cos\theta,
\end{equation}
where we neglect the meridional and vertical wind components \citep{ZhangEtAl2017a, PinoEtAl2022}.
The net Doppler shift is calculated by numerically integrating a single-line toy model in the visible hemisphere.
We assume the same Gaussian line profile with a $\sigma$ of 1 \kms (roughly the thermal broadening of Fe lines), but Doppler-shifted based on the $v_\mathrm{los}$ in each of the $192\times94$ pixelized patches on the planet's surface. 
At each orbital phase, we integrate the flux weighted by the projected surface area of each patch. The net Doppler shift is estimated from the peak location of the disc-integrated flux.
We then apply this evaluated Doppler shift to the 1D model spectrum at each orbital phase.

This integration also provides the broadening of line profiles as a function orbital phases. However, it does not match the observed phase-dependence of the FWHM, likely because of the simplified assumptions of no wind/turbulence dispersion and the identical line profile across different longitudes and latitudes. Therefore, we instead adopt the parametric form in Eq.~\ref{eq:line_width} for broadening our models. Future studies could improve the broadening prescription to better link it to the dynamical properties across the atmosphere.

\begin{figure*}[t]
\centering
    \includegraphics[width=\linewidth]{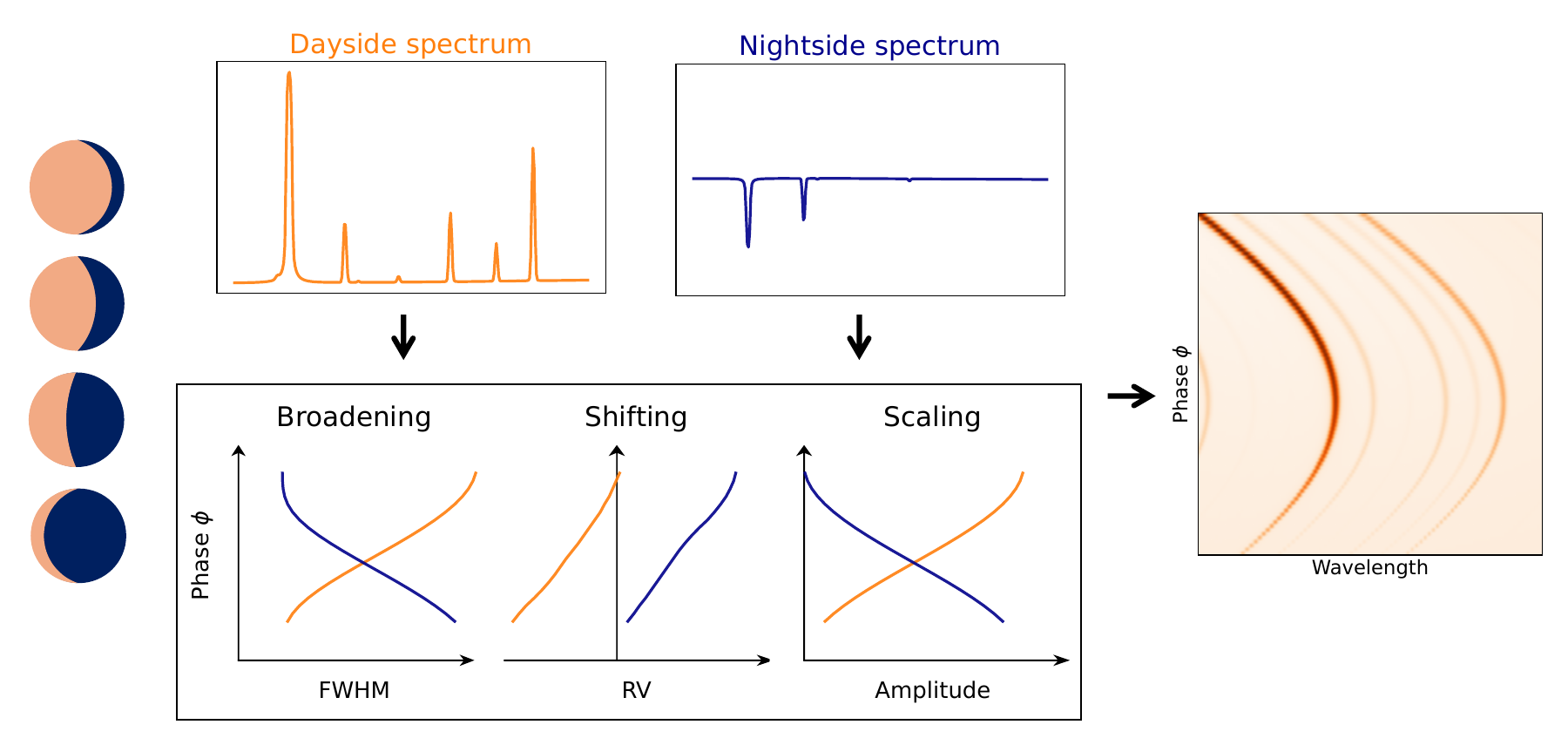}
    \caption{Illustration of the modeling framework for phase-resolved time-series spectra. With average dayside (thermal inversion; emission lines) and nightside (non-inverted profile; absorption lines) spectra, we apply phase-dependent scaling, broadening, and Doppler shifting to model the full time-series observations.
    \label{fig:schematic}}
\end{figure*}

With the atmospheric template as presented in Section~\ref{sec:model}, we apply the phase dependencies as described in Equations~\ref{eq:line_strength}, \ref{eq:line_width}, and \ref{eq:line_doppler}
by scaling, broadening, and shifting the template according to each orbital phase to generate the time-series model.
We account for the nightside emission by making a composite model that combines the dayside and nightside contributions. The nightside spectrum is also broadened, shifted, and scaled based on the opposite phase dependence ($0.5-\phi$) as illustrated in Fig.~\ref{fig:schematic}.
This allows us to place tight constraints on the nightside thermal properties in the case of non-detection.

We exclude the orbital phases of $\phi<0.11$ and $\phi>0.4$ from the retrieval analysis due to the stellar contamination as mentioned in Section~\ref{sec:analysis_ccf}.
We use data in all spectral orders except for the two reddest orders with wavelengths from 848 to 870 nm covering the Ca\,{\sc ii} triplet. 
We note that the LTE models are not adequate to reproduce the strong and broad line shape of the Ca\,{\sc ii} triplet, which will be discussed separately in Section~\ref{sec:result_retrieval}.

Then, we compare the model spectra to observations using the likelihood function from \cite{GibsonEtAl2020} \citep[see also][]{BrogiLine2019} as follows
\begin{equation}
    \ln \mathcal{L} = -\frac{N}{2}\ln \frac{\chi^2}{N}= -\frac{N}{2}\ln \left( \frac{1}{N} \sum_{i=1}^{N} \frac{(y_i - f_i)^2}{\sigma_i^2} \right),
\end{equation}
which folds in the optimized scaling term $\beta$ for observation uncertainties $\sigma_i$.

We assume uniform priors for the free parameters, including the temperature structure on dayside and nightside, separately ($\kappa_\mathrm{IR}$, $\gamma$, and $T_\mathrm{eq}$),
the orbital velocity \kp, the offset from systemic velocity $\Delta$\vsys~to account for the measurement uncertainty of \vsys, the maximum FWHM, and the wind speed $u_\varphi$.
We use the nested sampling tool \texttt{PyMultiNest} \citep{BuchnerEtAl2014}, 
which is a Python wrapper of the MultiNest method \citep{FerozEtAl2009} for the Bayesian inference. 
The retrievals are performed in importance nested sampling mode with a constant efficiency of 5\%. 
It uses 1000 live points to sample the parameter space and derives the posterior distribution of free 
parameters, as summarized in Table~\ref{tab:params}.

\begin{deluxetable}{ccc}
    \tablecaption{Priors and posteriors of \K~b retrievals. For the non-detected nightside emission, we include the 1$\sigma$ lower limits for $\kappa_\mathrm{IR}$ and $\gamma$.}
    \label{tab:params}
    \tablehead{\colhead{Parameter} & \colhead{Prior} & \colhead{Posterior} }
    \startdata
    \kp~(\kms) & $\mathcal{U}(200,300)$ & $ 244.8\pm 1.2 $ \\
    $\Delta$\vsys~(\kms) & $\mathcal{U}(-10, 10)$ & $  1.4\pm0.9  $ \\
    $u_\varphi$ (\kms) & $\mathcal{U}(0,20)$ & $  11.7 \pm 0.6 $ \\
    FWHM (\kms) &  $\mathcal{U}(5,20)$ & $  11.8\pm 0.5 $ \\
    $T_\mathrm{day}$ (K) & $\mathcal{U}(4000,5000)$ & $  4579\pm 120 $ \\
    $T_\mathrm{night}$ (K) & $\mathcal{U}(2000,3000)$ & - \\
    $\log \kappa_\mathrm{IR, day}$ (cm$^2\,$g$^{-1}$) & $\mathcal{U}(-5, -1)$ & $  -4.42\pm 0.11 $ \\
    $\log \kappa_\mathrm{IR, night}$  (cm$^2\,$g$^{-1}$)& $\mathcal{U}(-5, -1)$ &  $>-3.2$\\
    $\log \gamma_\mathrm{day}$ & $\mathcal{U}(0, 3)$ & $  0.61\pm0.05  $ \\
    $\log \gamma_\mathrm{night}$ & $\mathcal{U}(-3, 0)$ & $>-0.9$ \\
    \enddata
\end{deluxetable}

\section{Results} \label{sec:result}

\subsection{From phase-dependent Doppler shifts to winds}\label{sec:result_dynamics}

\begin{figure}[t]
\centering
    \includegraphics[width=\linewidth]{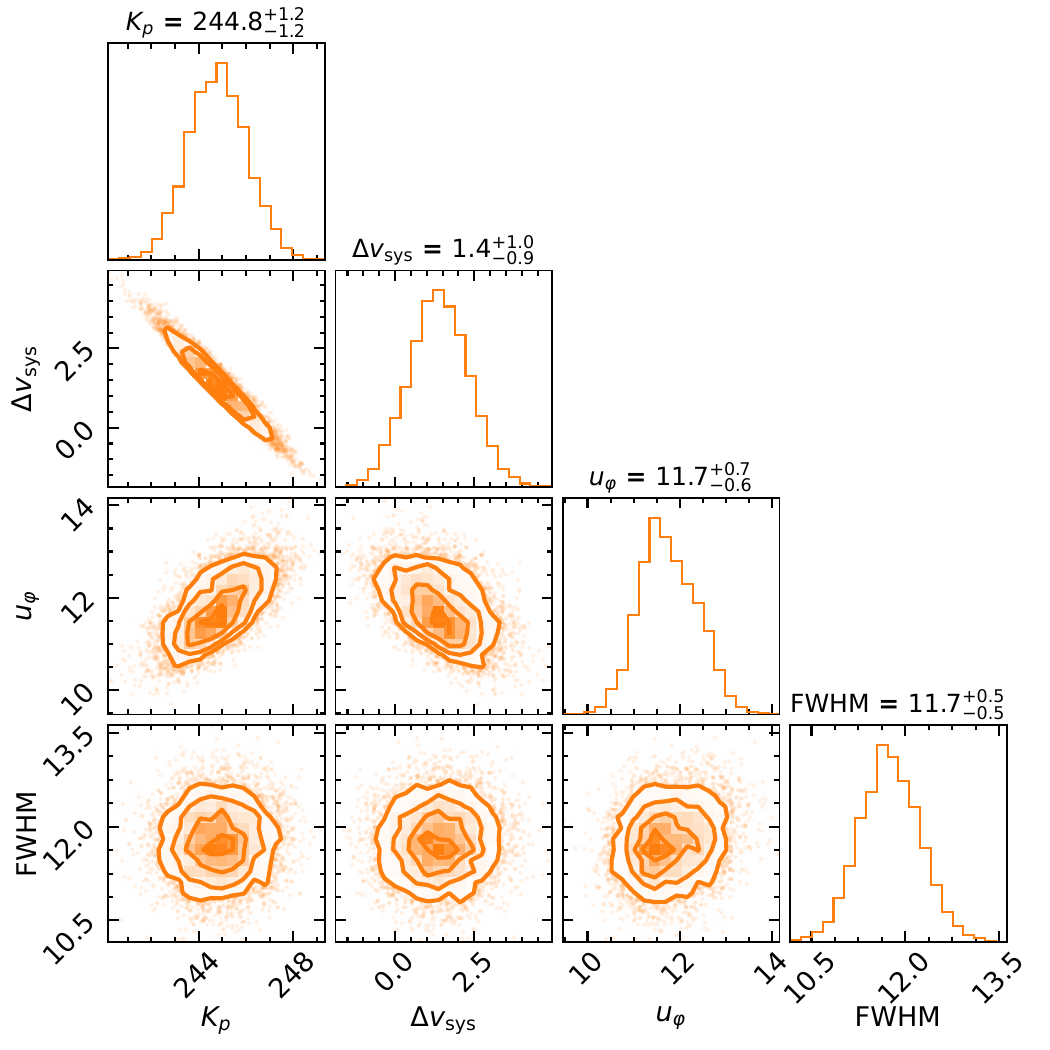}
    \caption{Corner plot of the retrieved orbital velocity \kp, systemic velocity offset $\Delta$\vsys, zonal wind speed $u_\varphi$, and line widths. 
    \label{fig:corner_v}}
\end{figure}

We present the CCF time series of \K~b observations in Fig.~\ref{fig:ccf_phase}.
To compute the CCFs, we use a static template (not broadened or shifted), as obtained from the best-fit model in the retrieval analysis 
(Section~\ref{sec:analysis_retrieval}), which is further discussed in Section~\ref{sec:result_retrieval}.
The CCFs show a significant planetary detection across almost all orbital phases. 
In particular, we detect emission from the crescent dayside at post-transit phases ($\phi=0.1-0.25$) for the first time.
This extends the phase-resolved detection of UHJ emission to post-transit phases.
In the star's rest frame, the planetary signal delineates a sinusoidal curve as expected from the orbital motion $K_\mathrm{p}\sin(2\pi\phi)$.
We then correct for the orbital motion and shift the CCFs to the planet's rest frame by assuming \kp=244.8 \kms, which is inferred from retrievals as discussed later in this section and shown in Fig.~\ref{fig:corner_v}.
In the planet's rest frame, the signal shows a progressive blueshift from $-0.4\pm1.0$ at $\phi\sim0.45$ to $-13.4\pm0.6$ \kms~at $\phi\sim0.1$.
We measure the net Doppler shift as a function of orbital phase by fitting a Gaussian profile to the average CCF within every phase bin of 0.02.
The phase dependence of the line widths, net Doppler shifts, and amplitudes are shown in Fig.~\ref{fig:phase_variation}.

The RV shifts as a function of orbital phases are driven by the planet's rotation plus atmospheric winds. 
As the tidally-locked planet rotates synchronously along with its orbital motion, it shows an increasing portion of the dayside hemisphere in view after the transit. 
This leads to a phase-dependent net Doppler shift of the disc integrated emission relative to the planet's rest frame because the flux is dominated by the dayside contribution.
If there is no atmospheric dynamics, the signal is expected to follow a sinusoidal curve with a maximum blueshift at the level of the spin rate \citep{ZhangEtAl2017a, PinoEtAl2022, WardenierEtAl2025}.
For \K~b, the tidally-locked spin rate corresponds to $v_\mathrm{eq} = 6.4 \pm 0.1$ \kms, which is not sufficient for explaining the RV shift of 13 \kms~observed at $\phi\sim0.1$.
Although any orbital eccentricity can also lead to the deviation of Doppler shifts from the assumed circular orbit, \K~b shows no evidence of an eccentric orbit based on the TESS light curves \citep[$e<0.007$ at 2$\sigma$,][]{WongEtAl2020}. This stringent upper limit on eccentricity corresponds to a maximum RV deviation of \kp$e\sim1.7$ \kms, which cannot account for the observed net Doppler shifts. We refer readers to \cite{PinoEtAl2022} for detailed discussion on the topic.
Overall, this unambiguously suggests the presence of atmospheric winds that add to the Doppler shift induced by planet rotation.

We test the impact of the uncertainties in ephemeris, \vsys~and \kp~on the measured net Doppler shifts.
For ephemeris, we propagate the uncertainties of the orbital period and transit midpoint \citep{KokoriEtAl2023} to the phase calculation, 
which results in a maximum uncertainty of 0.3 \kms~in RVs at individual orbital phases. This is much smaller than the resolution and line widths of the CCFs, and hence negligible.
For \vsys, the uncertainty only leads to a constant offset of the RVs, which does not affect the slope of the phase dependence as shown in Fig~\ref{fig:ccf_phase}.
However, the uncertainty of \kp~is critical in shaping the net RV trend across phases. 
The \kp~values of \K~b have large variations in the literature, therefore should be treated with caution.
As pointed out in \cite{HoeijmakersEtAl2024, WardenierEtAl2025}, emission spectroscopy analyses are expected to result in a reduced \kp~value 
compared to the true orbital velocity due to the unaccounted effect of planet rotation, which essentially counteracts the orbital motion. 
Thanks to the large phase coverage of our observations, we can include the effect from planet rotation and atmospheric winds (Equation~\ref{eq:line_doppler}) in retrievals 
and recover an accurate \kp~as well as wind speeds.
Using the parameterization described in Section~\ref{sec:analysis_retrieval}, we retrieve the \kp~and the
maximum zonal wind $u_\varphi$ from the phase-resolved observations, as shown in Fig.~\ref{fig:corner_v} and Table~\ref{tab:params}.
The marginalized values are $\Delta$\vsys=$1.4\pm0.9$ \kms, $K_\mathrm{p} = 244.8\pm 1.2$ \kms~and $u_\varphi = 11.7\pm 0.6$ \kms.
This leads to a \vsys~constraint of $-16.2\pm0.9$ \kms.
We note the negative correlation between \vsys~and \kp~because our observations only cover the half orbit prior to the eclipse. 
Combining the observations with post-eclipse phase coverage, which displays positive \vsys-\kp~correlation, will help to break the degeneracy and result in better constraints on these parameters. 

Using the updated \kp, we estimate the stellar mass following \cite{HoeijmakersEtAl2019}. We obtain a revised stellar mass of $M_\ast=2.26\pm0.03$~\Msun, stellar radius of $R_\ast = 2.28\pm0.03~R_\odot$, and planet radius of $R_p = 1.82\pm 0.03$~\Rjup, which uses the $\rho_\ast$ and $R_p/R_\ast$ values from \cite{GaudiEtAl2017}.
Combining the RV amplitude $K$ from \cite{PaiAsnodkarEtAl2022}, the planet mass is then $M_p = 2.22\pm 0.58$~\Mjup.

\begin{figure*}[t]
\centering
    \includegraphics[width=\linewidth]{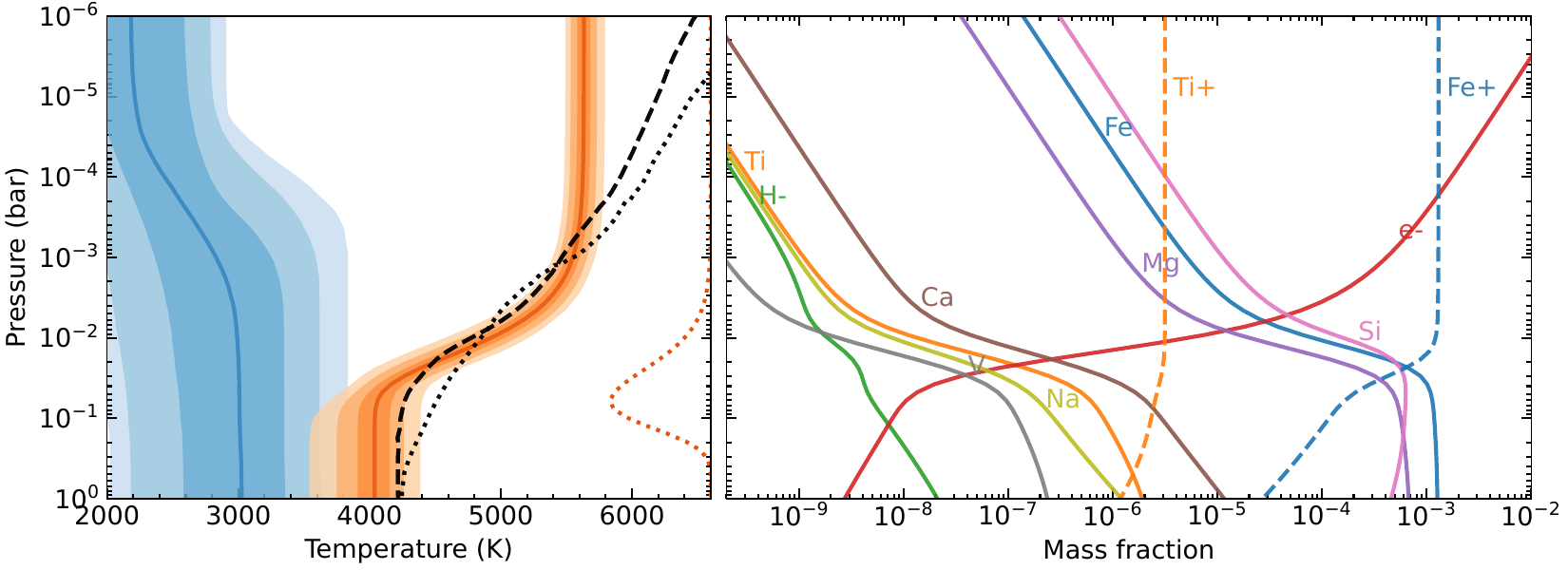}
    \caption{Left panel: Retrieved temperature-pressure profiles of the dayside (orange) and nightside (blue). The dotted orange line shows the wavelength-averaged emission contribution as a function of pressures. 
    The dashed and dotted black lines represent the self-consistent T-P profiles from \cite{LothringerEtAl2018, FossatiEtAl2021} for comparison.
    Right panel: equilibrium chemical abundance profiles for the mean dayside T-P structure.
    \label{fig:pt-chem}}
\end{figure*}

\subsection{Thermal Structure and Chemistry} \label{sec:result_retrieval}

The retrieval analysis also provides constraints on the thermal structure of \K~b. 
The retrieved P-T profile, with a strong temperature inversion in the dayside, is shown in Fig.~\ref{fig:pt-chem}.
It agrees well with self-consistent models for \K~b \citep{LothringerEtAl2018, FossatiEtAl2021} in the pressure regimes probed by emission lines
($P \sim 10^{-4}-10^{-1}$ bar, as shown by the emission contribution in Fig.~\ref{fig:pt-chem}).
We find similar results for pressures higher than $10^{-4}$ bar, while the lower pressure regions are not well constrained by the observations.
In addition, NLTE effects are expected to be significant in the upper atmospheres, leading to unreliable results with our LTE retrievals.

According to phase curve measurements, the nightside temperature of \K~b is estimated to be $\sim 2500-3000$ K \citep{MansfieldEtAl2020, WongEtAl2020, JonesEtAl2022}.
This temperature is high enough to keep the metal species in gaseous form, and hence we expect the nightside to show absorption spectral features (because of non-inverted P-T profiles) that the high-resolution observations are still sensitive to.
However, we obtain a non-detection of the nightside emission with our KPF observations.
Our retrieval with a composite day- and night-side model places an upper limit on the nightside thermal structure as shown in Fig~\ref{fig:pt-chem}.
This non-detection implies a shallow vertical temperature gradient on the nightside.
We can rule out large temperature gradients as suggested by the $\gamma$ parameter (Table~\ref{tab:params}).
We plot an example of the fitted nightside spectrum model in Fig.~\ref{fig:template} for a visual comparison of the line strengths to those of the dayside emission spectrum. It suggests that the nightside line strength is over an order of magnitude smaller than the dayside signal, again highlighting the challenge of detecting nightside emission \citep{WardenierEtAl2025}.
Our temperature constraint is consistent with the nightside T-P profiles from the GCMs presented in \cite{MansfieldEtAl2020}. 
However, the observations are not sensitive enough to distinguish the nightside thermal profiles predicted for different levels of atmospheric drag. 
A weak drag would lead to efficient heat redistribution and therefore relatively isothermal T-P profiles on the nightside, while a strong drag means that the nightside is dominated by radiative processes and retains a cooler upper atmosphere.
Upcoming JWST phase curve observations will provide better insights into the nightside structure \citep{StevensonEtAl2025}.

\begin{figure*}[t]
\centering
    \includegraphics[width=\linewidth]{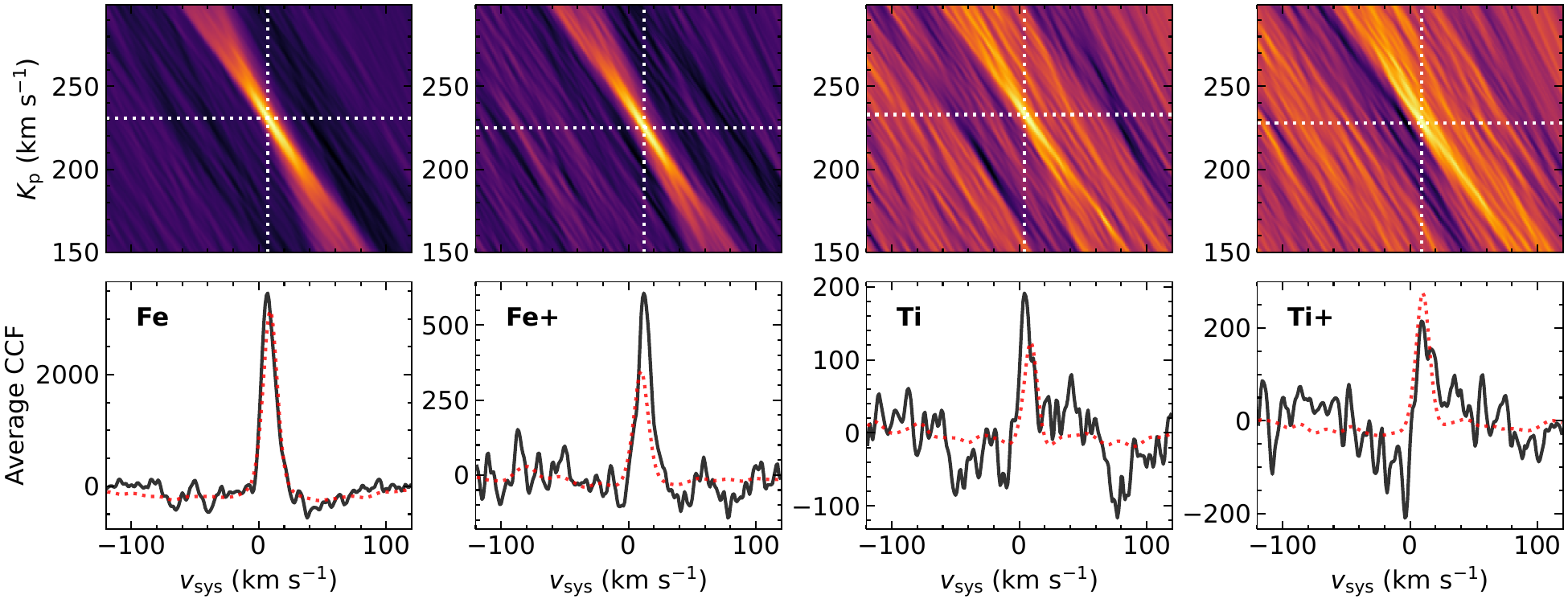}
    \includegraphics[width=\linewidth]{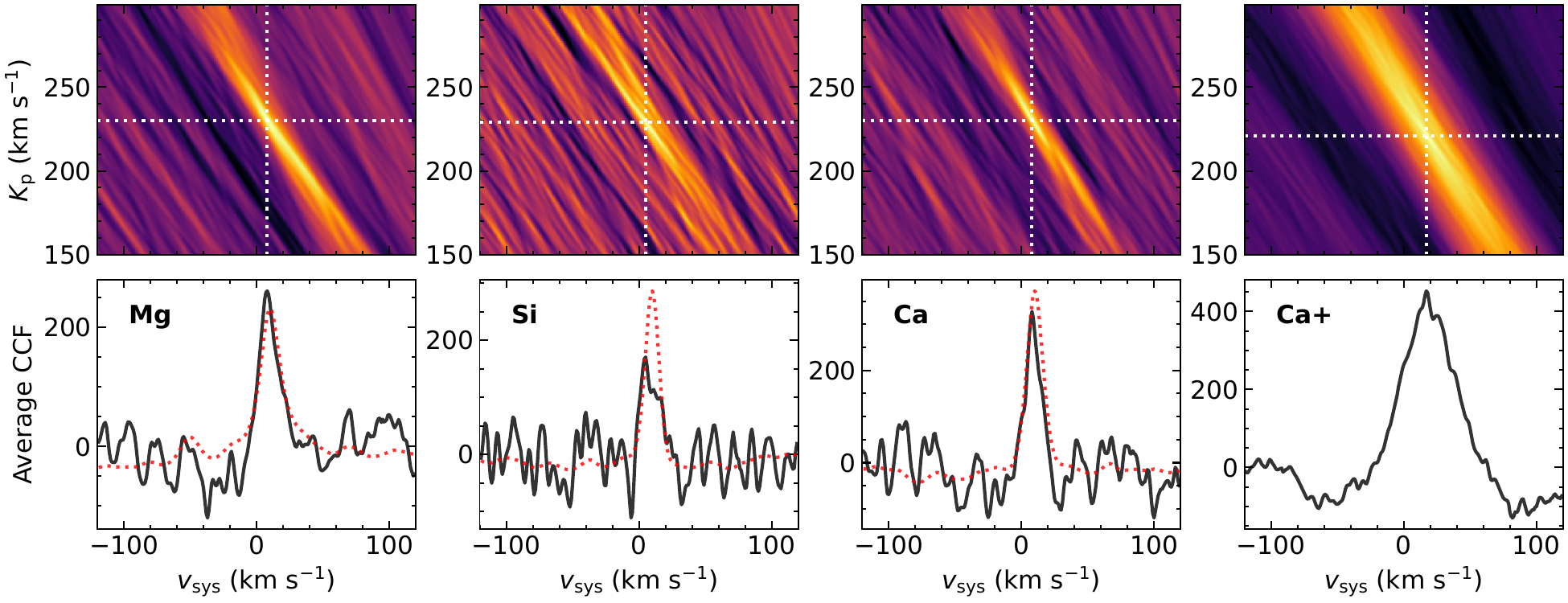}
    \caption{\vsys-\kp~diagram and average 1D CCF detection of Fe\,{\sc i}, Fe\,{\sc ii}, Ti\,{\sc i}, Ti\,{\sc ii}, Mg\,{\sc i}, Si\,{\sc i}, Ca\,{\sc i} and Ca\,{\sc ii}. The white dotted lines mark the locations of CCF peaks. The red dotted lines show the model prediction of CCF signals.
    \label{fig:kp_vsys1}}
\end{figure*}


Regarding the chemical composition, we detect emission from 
Fe\,{\sc i}, Fe\,{\sc ii}, Ti\,{\sc i}, Ti\,{\sc ii}, Ca\,{\sc i}, Ca\,{\sc ii}, Mg\,{\sc i}, Si\,{\sc i} on the dayside. 
The mass fractions of these species in chemical equilibrium are shown in Fig.~\ref{fig:pt-chem}.
The onset of thermal ionization of neutral metals occurs near the photosphere ($P\sim 10^{-2}-10^{-1}$ bar) and dominates the chemistry at lower pressures.
We present the CCF detections for individual species in Fig.~\ref{fig:kp_vsys1}. 
The template of the individual species used in the cross-correlation is built from the retrieved best-fit model minus the contribution from all other species except the one being analyzed, as shown in Fig.\ref{fig:template}.
To assess the observational deviation from the chemical-equilibrium model, we compare the observed average CCFs with the modeled CCFs, 
as shown in the bottom panels in Fig.~\ref{fig:kp_vsys1}. The model shows generally good agreement with the observations, while underestimating the emission strength of Fe\,{\sc ii} and Ti\,{\sc i}, and overestimating the Si\,{\sc i} signal.
Similar underestimation of Fe\,{\sc ii} signal has been reported in transmission studies as well \citep{HoeijmakersEtAl2019}.
The discrepancy is likely caused by NLTE effects, the non-equilibrium chemistry (e.g., photoionization), and the non-hydrostatic structure in the upper atmosphere.

We explore the potential velocity discrepancy across different species in Fig.~\ref{fig:kp_vsys1}. 
The CCF peak locations are consistent for all neutral species, while the peak for ionized species tends to shift to a lower \kp~and higher \vsys.
To quantify the velocity difference, we carry out a retrieval with separated wind speed parameters for neutral (e.g., Fe\,{\sc i}, Ti\,{\sc i}, etc.) and ionized species (Fe\,{\sc ii} and Ti\,{\sc ii}), respectively. We find consistent values for the two retrieved $u_\varphi$, therefore no clear evidence of different dynamical behavior for ions versus neutrals. 
Instead, the differential velocity shifts are expected to be more prominent in atomic gases (e.g., H\,{\sc i}, Na\,{\sc i}, and Ca\,{\sc ii}) that probe distinct pressure levels or molecular species (e.g., \HTWOO, and OH) that show spatial gradients due to dissociation \citep{KesseliEtAl2022, BrogiEtAl2023, WardenierEtAl2024, SeidelEtAl2025}.

\begin{figure*}[t]
\centering
    \includegraphics[width=\linewidth]{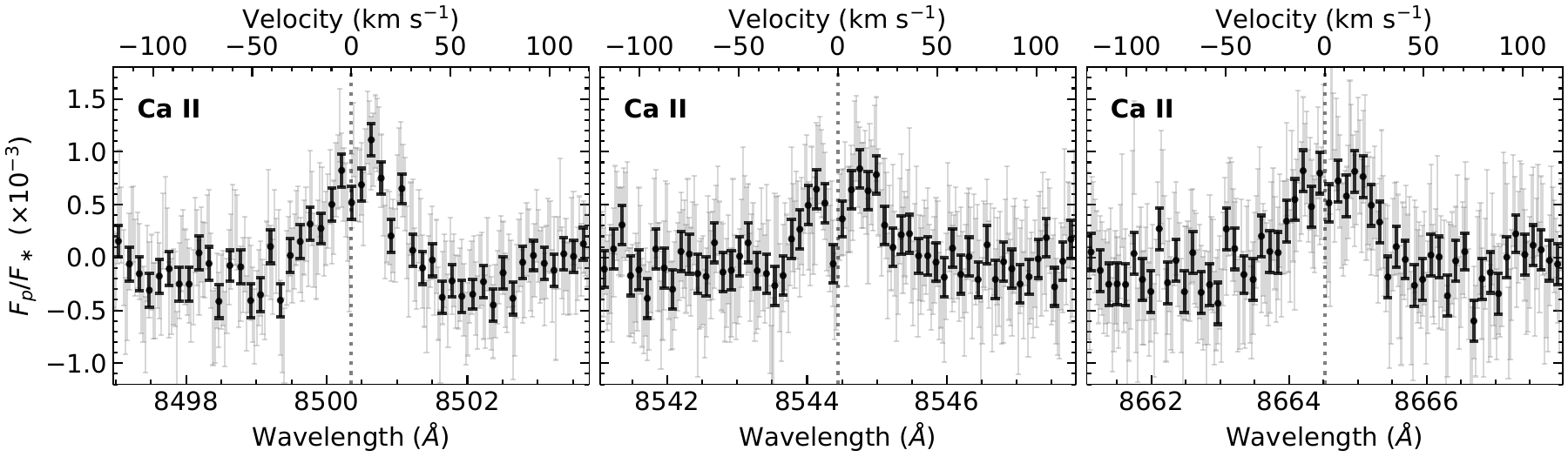}
    \caption{Resolved emission from Ca\,{\sc ii} triplet lines at 8500, 8544, and 8564 \AA~in \K~b. The black data points represent the binned data in every 5 pixels of the original data (in gray).
    \label{fig:CaII}}
\end{figure*}

We resolve individual lines of the Ca\,{\sc ii} triplet at 8500.3, 8544.4, and 8664.5 \AA, as shown in Fig.~\ref{fig:CaII}.
The broadening of the triplet (FWHM$\sim 45$ \kms) is significantly larger than that of other metal lines, indicating a distinct thermal and dynamical regime \citep{ZhangEtAl2022}. 
We note that the wavelengths near the Ca\,{\sc ii} lines are excluded from the aforementioned retrieval analysis because the lines originate from a higher 
region of the atmosphere where NLTE and hydrodynamical effects play significant roles and cannot be readily modeled with our framework.
We encourage future studies to perform NLTE modeling on the Ca\,{\sc ii} triplet combining emission with transmission observations \citep{YanEtAl2019, BorsatoEtAl2023, DArpaEtAl2024},
which are particularly constraining on the thermal structure and dynamics in the upper atmosphere.

\section{Discussion} \label{sec:discussion}

\subsection{Comparison to GCMs} \label{sec:gcm}

To investigate the implications of the phase-dependent Doppler shifts on atmospheric dynamics, we compare our observations to the predictions from GCMs \citep{TanKomacek2019} for \K~b as presented in \cite{MansfieldEtAl2020}.
The model uses the double-gray approximation in the radiative transfer calculation with wavelength-independent opacities within the visible and infrared band.
As the double-gray assumption does not apply at low pressures because the solution converges to an isotherm \citep{RauscherMenou2012}, the model top boundary was cut at 1 mbar. 
This pressure range is sufficient because our emission observations mainly probe pressures of $10^{-1}-10^{-3}$ bar as shown in Fig~\ref{fig:pt-chem}.
The GCMs include the cooling and heating effects due to dissociation and recombination of molecular hydrogen, which are suggested to be critical for heat redistribution in UHJs \citep{BellCowan2018}. 
We use multiple GCMs with different levels of frictional drag, parameterized by a single drag timescale $\tau_\mathrm{drag}$, which is used to represent the effects of magnetic field and turbulence in the linear regime. 
We consider $\tau_\mathrm{drag}=10^3$ and $10^7$ s, ranging from strong to weak drag. 
The simulations predict various strengths of day-to-night flows at $10^{-2}$ bar subject to the atmospheric drag. The weak drag can result in locally supersonic 
wind speeds up to 15 \kms, while the strong drag can slow down the wind speed to 3 \kms. 

\begin{figure}[t]
\centering
    \includegraphics[width=\linewidth]{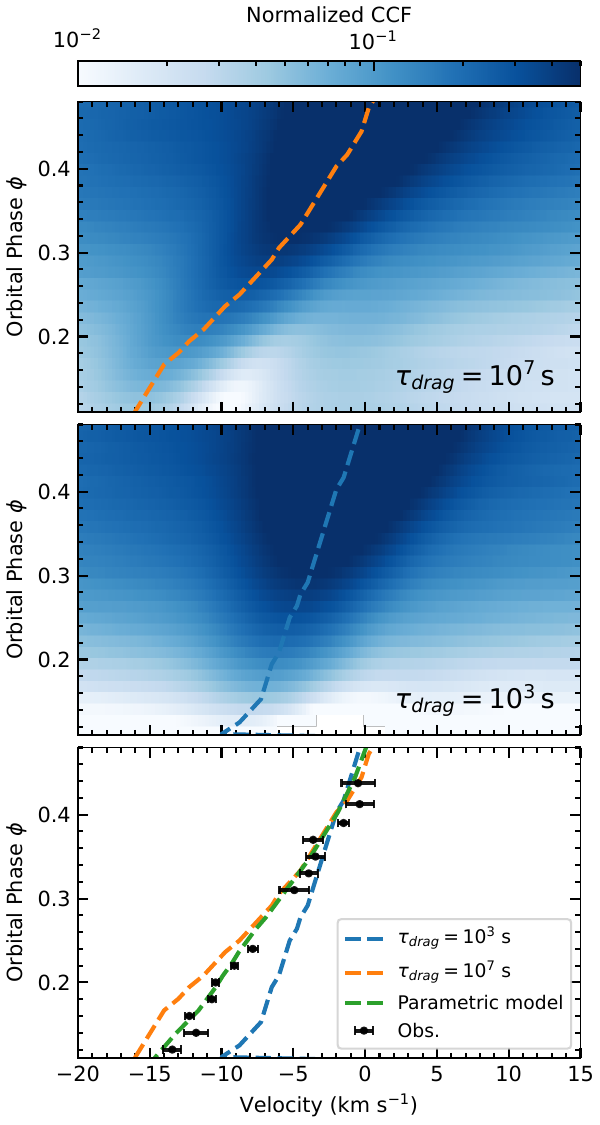}
    \caption{
    Net Doppler shifts of the planetary emission as a function of orbital phases compared to GCM predictions.
    Top and middle: predicted CCF maps for GCMs with different drag timescales: weak drag $\tau_\mathrm{drag}=10^7\,$s and strong drag $\tau_\mathrm{drag}=10^3\,$s. The dashed lines mark the location of CCF peak at each orbital phase.
    Bottom: comparison of the phase-velocity trend between the observations and GCMs. The black data points are the measured velocities of the line center as also shown in Fig.~\ref{fig:phase_variation}. 
    The observations prefer the GCM with $\tau_\mathrm{drag}=10^7$ s.
    The green dashed line shows the retrieved best-fit wind model with $u_\varphi = 11.7$ \kms.
    \label{fig:gcm_ccf}}
\end{figure}

We postprocess the outputs of the GCMs with the 3D Monte-Carlo radiative transfer code gCMCRT \citep{LeeEtAl2022} to simulate emission spectra at a range of orbital phases from 0 to 0.5. We refer readers to \cite{WardenierEtAl2025} for a detailed description of the code. 
We obtain a static dayside template by zeroing out any dynamical effects, and then cross-correlate it with the simulated time-series spectra. The CCF maps as a function of orbital phase are shown in Fig.~\ref{fig:gcm_ccf}. 
We note the similarly slanted traces in the simulated CCFs and the observations (Fig.~\ref{fig:ccf_phase}), which is driven by planet rotation and winds \citep{ZhangEtAl2017a, WardenierEtAl2025}.
Various levels of atmospheric drag can affect the wind pattern, therefore resulting in a different phase dependence of the net Doppler shift. Fig.~\ref{fig:gcm_ccf} shows the phase variation of the CCF peak velocities for different drag timescales. The extracted net RVs from observations are also shown for comparison. 
The observations match the $\tau_\mathrm{drag}=10^7 \ \mathrm{s}$ model best. The strong day-to-night wind predicted in this weak drag model is necessary to account for the large blueshifts ($\sim13$ \kms) seen in the observations.

Although the phase dependence of the net Doppler shifts can be generally reproduced by the GCMs, we caution that inferring accurate atmospheric drag from these models remains difficult.
As a result of the double-gray approximation and the limited temperature range of opacity tables, these GCMs are unable to match the strong dayside thermal inversion from our retrievals. It is unclear how the underestimated dayside temperature would affect the wind structure. Future 3D modeling studies are needed to reassess the approximations and produce more realistic predictions. Our high-resolution phase curve observations serve as valuable benchmarks for GCMs.

Inspecting the GCM results, we find that the zonal winds across the dayside hemisphere can be well described by a sinusoidal function of longitude as shown in Fig.~\ref{fig:u_phi}.
This supports our parameterization of the winds in Equation~\ref{eq:line_doppler}, allowing us to retrieve meaningful wind speeds directly from the observations.
Independent from the GCMs, our retrieved wind speed of $u_\varphi = 11.7\pm 0.6$ \kms~is consistent with the GCM prediction with a drag timescale of $\tau_\mathrm{drag} \sim 10^7$ s.
The day-to-night flow becomes supersonic near the terminator regions, leading to blueshifts up to $\sim-10$ \kms~as also observed in transmission spectroscopy \citep{AsnodkarEtAl2022, DArpaEtAl2024, StangretEtAl2024}. 
In parallel, phase curve measurements suggest a relatively small day-to-night temperature contrast, and therefore an efficient heat redistribution in \K~b, which can be attributed to molecular hydrogen dissociation and recombination \citep{BellCowan2018, MansfieldEtAl2020, WongEtAl2020}. 
Both the thermal phase curve and the high-resolution Doppler measurements indicate a consistent picture of weak atmospheric drag, vigorous circulation, and strong day-to-night flows in \K~b.

Supersonic winds are commonly predicted by models solving for hydrostatic primitive equations and filtering out sound waves, as presented here. 
More complex, hydrodynamic Navier-Stokes models that allow for shocks also predict supersonic equatorial jets in hot Jupiters, despite showing $\sim15\%$ velocity fluctuations due to the Kelvin-Helmholtz instability \citep{FromangEtAl2016}. Therefore, it is possible that \K~b can maintain locally supersonic winds and display some variability in velocity. However, future modeling studies in the ultra hot regime of \K~b are needed to understand the potential differences between hydrostatic and hydrodynamic models under such extreme conditions.
The excellent data quality in this paper provides the best opportunity to date to benchmark these GCMs.

\subsection{Magnetic fields}

Constraints on the atmospheric drag can be useful for understanding fundamental properties of UHJs that drive the atmospheric dynamics. 
In our GCMs, the atmospheric drag is parameterized as a Rayleigh drag force, characterized by a single timescale $\tau_\mathrm{drag}$ throughout the planet.
Under this assumption, we can link the drag timescale to the magnetic field strength, following \cite{PernaEtAl2010, ArcangeliEtAl2019}:
\begin{equation}
    \tau_\mathrm{drag} \sim \frac{4\pi\rho\eta}{B^2cos\theta},
\end{equation}
where $\rho$ is the density of the atmosphere, $\eta$ is the magnetic diffusivity related to the ionization fraction ($n_e$), $B$ is the magnetic field strength, 
and $\theta$ is the angle between the magnetic field and the flow direction.
We estimate that a drag timescale of $\tau_\mathrm{drag} \sim 10^7$ s corresponds to a dipole magnetic field strength of $B_\mathrm{dip} \sim 3 \times 10^{-3}$ G, assuming a maximum efficiency ($\theta=0$). 
This does not account for the direction dependence and spatially inhomogeneous strengths of the magnetic drag \citep{RauscherMenou2012, BatyginEtAl2013}. 

There seems to be a slight tension between the inferred weak magnetic drag with the expected strong B fields in this highly irradiated atmosphere.
However, we caution that the simplified assumptions are likely invalid for the extreme atmosphere of \K~b. 
The Rayleigh drag assumption only holds in the limit of small magnetic Reynolds number $R_m = u_\varphi H /\eta < 1$, 
where $u_\phi$ is the zonal wind speed and $H$ is the scale height. 
We estimate \K~b to be in the high magnetic Reynolds number regime of $R_m \sim 10^5$.
In this non-linear regime, the magnetic field can change the strength and direction of Lorentz force in a time-dependent manner, leading to either acceleration or deceleration of winds \citep{RogersKomacek2014, HindleEtAl2021a, DietrichEtAl2022}.
Therefore, magnetohydrodynamic (MHD) models are required to better assess the impact of magnetic fields on the atmospheric dynamics \citep{BeltzEtAl2022, BeltzRauscher2024, KesseliEtAl2024}. However, this has been challenging for the ultra hot regime of \K~b due to numerical limitations.

Alternatively, \cite{SavelEtAl2024a} proposes to constrain the magnetic field strength using differential wind velocities between heavy ions and neutral atoms in transmission, as the ionized gas is more strongly coupled to the magnetic field. 
However, emission observations probe higher pressure levels ($>$1 mbar) where the ion–neutral velocity difference due to magnetic effects is expected to be negligible \citep{SavelEtAl2024a}. 
This is supported by the consistent winds inferred from neutral and ionized species, as discussed in Section~\ref{sec:result_retrieval}.

\subsection{Atmospheric winds in the UHJ population}

Most previous studies of hot Jupiters have used high-resolution transmission spectroscopy to constrain atmospheric winds, and have found that blueshifted absorption features appear to be ubiquitous in these observations \citep[see the review in ][]{Snellen2025}.
WASP-76 b and WASP-121 b represent the most favorable transmission targets with phase-resolved observations, displaying significant blueshifted atomic absorptions of up to 10 \kms~towards the ends of the transits \citep{EhrenreichEtAl2020, KesseliSnellen2021, BorsaEtAl2021a}. 
These blueshifts correspond to day-to-night wind speeds of $\sim6.9$ \kms~for WASP-76 b and 5.6 \kms~for WASP-121 b, as retrieved by \cite{GandhiEtAl2023a}.
WASP-172 b is inferred to show winds up to $10-14$ \kms~based on the blueshifts of the Na doublet lines \citep{SeidelEtAl2023a}.
Moreover, different atoms and molecules tracing different pressures or longitudes exhibit distinct phase dependencies in transmission signals \citep{WardenierEtAl2024, SeidelEtAl2025}, unlocking information on the 3D structures.
In addition, dayside measurements also probe dynamical signatures. Using the redshifted dayside emission of CO and Fe, \cite{LesjakEtAl2025} constrains a day-to-night wind velocity of $\sim4.4$ \kms~for WASP-189 b.
In contrast, \cite{CostaSilvaEtAl2024} reports blueshifted ($-4.7 \pm 0.3$ \kms) Fe~{\sc i} emission on the dayside of WASP-76 b and proposes that this can be caused by the material upwelling from the hotspot. 
In contrast, a recent study by \cite{GuilluyEtAl2025} contests the previous measurements and finds no blueshift along the \vsys~axis but a negative shift in \kp, in line with the predictions from \cite{WardenierEtAl2025}. 
Comparing winds in different UHJs has been challenging due to the inhomogeneity of these diverse measurements.

For many UHJs, this picture is additionally  complicated by their fast-rotating early-type host stars, which make it difficult to accurately measure \kp~and \vsys, hindering robust constraints on atmospheric dynamics. 
Different treatments of \kp~in fitting can lead to contrary conclusions on the presence of winds \citep{PrinothEtAl2022, PrinothEtAl2023, PrinothEtAl2024}. 
For atomic species, the stellar pulsation, Rossiter-McLaughlin effect, and center-to-limb variation can introduce contamination to the planetary transmission signals, making this approach less effective for many UHJs. 

Our high-resolution phase curve study suggests an alternative way of measuring winds that circumvents these systematics.
The resolved phase-dependence of the emission line profiles helps to break the degeneracy between orbital motion and atmospheric dynamics, as we essentially rely on the relative Doppler shifts across a wide range of phases, which can be robustly measured despite the uncertainties in absolute values.
In comparison to high-resolution transmission spectroscopy, emission observations trace the lower atmospheric region, where the hydrostatic and LTE assumptions are more valid, therefore making them better for benchmarking GCMs and revealing the underlying physics. 

Despite these uncertainties, our new observations strongly suggest that \K~b has the strongest day-to-night winds ($\sim$12 \kms) measured for an UHJ to date, consistent with its uniquely high irradiation. 
It is interesting to put this measurement into context at the population level.
Do hotter planets always develop stronger winds? How does the wind speed correlate with planet properties such as irradiation temperature, day-to-night contrast, heat recirculation efficiency, and magnetic strength?
These questions call for extending such studies to more UHJs using the suite of high-resolution instruments, such as Keck/KPF \citep{GibsonEtAl2020a}, Keck/HISPEC \citep{MawetEtAl2019}, VLT/ESPRESSO \citep{PepeEtAl2021}, VLT/CRIRES+ \citep{DornEtAl2023}, GEMINI/MAROON-X \citep{SeifahrtEtAl2018}, GEMINI/IGRINS-2 \citep{OhEtAl2024}, etc.
Combined with space-based phase curve measurements \citep{SingEtAl2025}, these observations will soon unveil trends in atmospheric dynamics with planet properties and constrain the driving mechanisms behind these trends in UHJs.
This will in turn open up a novel window into the wind patterns of canonical hot Jupiters in the era of ELTs.

\section{Conclusion} \label{sec:conclusion}

We characterize the atmosphere of the ultra-hot Jupiter \K~b using phase-resolved emission spectroscopy with KPF ($\mathcal{R}\sim98\,000$) in optical wavelengths. The observations cover a wide range of orbital phases from 0.05 to 0.48 and result in detection of planetary dayside emission throughout the half orbit from post-transit to pre-eclipse. Our findings are summarized as follows.

\begin{itemize}
\item We resolve planetary emission at post-transit phases ($\phi=0.1-0.25$) under high spectral resolution for the first time. 
\item The temporally resolved emission signals show phase-dependent profiles in terms of line strengths, widths, and Doppler shifts. The net Doppler shift varies progressively from $-13.4\pm0.6$ to $-0.4\pm1.0$ \kms, unambiguously driven by planet rotation plus atmospheric winds.
\item We parameterize the phase variations of line profiles and perform retrieval analysis on the full dataset of the high-resolution phase curve observations. 
We retrieve the thermally inverted dayside vertical temperature profile and place an upper limit on the nightside thermal structure. 
\item The retrieval reveals a supersonic day-to-night wind speed up to $11.7\pm0.6$ \kms~on on the emerging dayside of \K~b, representing the most extreme atmospheric winds in hot Jupiters to date.
\item We report cross-correlation detection of atomic species, including Fe\,{\sc i}, Fe\,{\sc ii}, Ti\,{\sc i}, Ti\,{\sc ii}, Ca\,{\sc i}, Ca\,{\sc ii}, Mg\,{\sc i}, Si\,{\sc i}, on the dayside.
\item We run 3D radiative transfer on double-gray GCM outputs to simulate the phase variation of the high-resolution emission spectra.
The comparison to GCMs points towards a weak atmospheric drag, consistent with the retrieved strong wind speed ($\sim12$ \kms). 
This also agrees with the photometric phase curves, which suggest high temperature ($\sim3000$ K) on the nightside and relative efficient heat redistribution. 
\item More realistic GCMs are required in the extreme parameter regime of \K~b to explore the full implications of the observations on atmospheric drag and magnetic fields.
\end{itemize}

We highlight the power of such high-resolution phase curve observations in directly probing atmospheric winds, thermal, and chemical structures in hot Jupiters.
The presented \K~b data are invaluable for benchmarking 3D circulation models and developing new retrieval techniques to map out the winds.
With the increasing sample of UHJs characterized by high-resolution transmission and emission spectroscopy, the emerging population-level trend will allow us to answer fundamental questions on circulations, climates, and magnetic fields of highly irradiated exoplanets.

\begin{acknowledgements}

    We thank the KPF Community Cadence observing team for their support in the observations.
    We thank the referee and editors for their constructive comments. We thank Ignas Snellen for helpful discussion.
    Y.Z. acknowledges the support from the Heising-Simons Foundation 51 Pegasi b Fellowship (grant \#2023-4298).
    The computation was carried out on the Caltech High-Performance Cluster.
    J.P.W. acknowledges support from the Trottier Family Foundation via the Trottier Postdoctoral Fellowship, as well as support from the Canadian Space Agency (CSA) under grant 24JWGO3A-03.
    A.H. is supported by the National Science Foundation Graduate Research Fellowship under Grant No. 2141064 and the MIT Dean of Science Fellowship.
    L.P. acknowledges support from the INAF Mini-Grant 2023 “Atmospheric structure, dynamics, and composition of hot gas giant exoplanets with high dispersion emission spectroscopy” (PI: L. Pino).
    
\end{acknowledgements}

\vspace{5mm}
\facilities{Keck:I/KPF}

\software{
\texttt{numpy}~\citep{HarrisEtAl2020},
\texttt{scipy}~\citep{VirtanenEtAl2020},
\texttt{matplotlib}~\citep{Hunter2007},
\texttt{astropy}~\citep{astropy:2013, astropy:2018, astropy:2022},
\texttt{petitRADTRANS}~\citep{MolliereEtAl2019}, 
\texttt{PyMultiNest} \citep{BuchnerEtAl2014},
\texttt{corner}~\citep{Foreman-Mackey2016}
}

\vspace{5mm}

\appendix

\section{The KELT-9 system} \label{app:system}

\K~b is the hottest exoplanet (\Teq $\sim4000$ K) discovered to date \citep{GaudiEtAl2017}. It has a 1.48-day, nearly polar orbit around an A0 star \citep{GaudiEtAl2017, AhlersEtAl2020}. 
At such a high temperature, its atmospheric chemistry resembles a dwarf star, with neutral and ionized metal species, accessible with high-resolution spectroscopy in the optical \citep{KitzmannEtAl2018, LothringerEtAl2018}. 
\K~b features the largest number of atomic species detected in an exoplanet atmosphere to date, such as H\,{\sc i}, Fe\,{\sc i}, Fe\,{\sc ii}, Ti\,{\sc i}, Ti\,{\sc ii}, Ca\,{\sc i}, Ca\,{\sc ii}, Mg\,{\sc i}, Si\,{\sc i}, O\,{\sc i}, and many more \citep{HoeijmakersEtAl2018a, YanHenning2018, CauleyEtAl2019, YanEtAl2019, TurnerEtAl2020, WyttenbachEtAl2020, BorsaEtAl2021, LangeveldEtAl2022, Bello-ArufeEtAl2022, AsnodkarEtAl2022, Sanchez-LopezEtAl2022, BorsatoEtAl2023, LowsonEtAl2023, BorsatoEtAl2024, DArpaEtAl2024, StangretEtAl2024}.

The presence of day-to-night winds in \K~b was under debate in the literature. 
\cite{HoeijmakersEtAl2019} reports no blueshifts of the transmission signals, while \cite{AsnodkarEtAl2022} updates the ephemeris and reports blueshifts due to winds up to $\sim10$ \kms~\citep[see also ][]{DArpaEtAl2024, StangretEtAl2024}.
The discrepancy mainly originates from the choice of ephemeris, the instrument RV zero-points, and uncertainties in the systemic and orbital velocities. 
As a result of the host star's rapid rotation \citep[$v\sin i\sim$ 115 \kms,][]{KamaEtAl2023}, the orbital properties and planet mass are challenging to constrain from high-resolution spectroscopy and radial velocity (RV) measurements. 
Various studies, using different instruments, spectral lines, or analysis methods, report different values - the systemic velocity \vsys~varies from $-21.6\pm 0.8$ to $-17.7\pm0.1$ \kms; the orbital velocity \kp~ranges from $234 \pm 1$ to $269\pm6$ \kms; the stellar mass ranges from $1.98\pm 0.02$ to $2.32\pm 0.16$ \Msun; the planet mass ranges from $2.17 \pm 0.56$ to $2.88 \pm 0.35$ \Mjup~in the literature \citep{BorsaEtAl2019, HoeijmakersEtAl2019, PaiAsnodkarEtAl2022}.
These uncertainties limit our ability to characterize the subtle effects from atmospheric dynamics.

In addition to absorption in transmission, Fe\,{\sc i}, Fe\,{\sc ii}, Ti\,{\sc i}, Ti\,{\sc ii}, Ca\,{\sc ii}, Mg\,{\sc i}, and Si\,{\sc i} emission from the dayside has been detected, confirming the inverted temperature structure \citep{PinoEtAl2020, KasperEtAl2021, PinoEtAl2022, Ridden-HarperEtAl2023}. 
Secondary eclipse observations using the Hubble Space Telescope's Wide Field Camera 3 
show a turnoff at 1.4 $\mu$m that has been attributed to either TiO, VO and FeH \citep{ChangeatEdwards2021}, or to strong absorption by H$^-$ that points to a high metallicity or a high ionization fraction whose origin remain elusive \citep{JacobsEtAl2022}.
However, ground-based observations find no evidence of these molecules \citep{KesseliEtAl2020, KasperEtAl2021, HayashiEtAl2024}. 
Phase curve observations from the space, including Spitzer \citep{MansfieldEtAl2020}, TESS \citep{WongEtAl2020}, and CHEOPS \citep{JonesEtAl2022}, suggest brightness temperatures at the dayside and nightside of $\sim 4500$ and $\sim 3000$K, respectively. This relatively shallow temperature contrast indicates a high day-to-night heat redistribution efficiency of $\epsilon \sim 0.3-0.5$, in comparison to typical UHJs with $\epsilon < 0.1$ \citep{DangEtAl2024}.
The dissociation and recombination of molecular hydrogen has been invoked to explain the enhanced energy transport \citep{BellCowan2018, MansfieldEtAl2020}. 
Using high-resolution emission spectroscopy, \cite{PinoEtAl2022} finds a deviation of the planetary RV from a circular orbit, likely caused by day-to-night atmospheric winds. 
Such studies with large phase coverage provide an avenue to break degeneracies between the orbital motion and atmospheric winds in transmission observations.

\section{Telluric correction}\label{app:molecfit}

We use the ESO sky tool \texttt{molecfit} to correct for telluric features in individual exposures.
It uses a line-by-line radiative transfer model (\texttt{LBLRTM}) to derive telluric transmission spectra that can best fit the observations.
The fitting parameters include the molecular abundances of telluric \HTWOO~and O$_2$, the instrument resolution (modeled with a Gaussian kernel), the continuum level of each spectral order (modeled by third-order polynomials).
We perform fitting in the following wavelength regions: [6867, 6933], [7185, 7258], [7570, 7720], and [8137, 8348] \AA, covering the main O$_2$ and \HTWOO~bands.
The wavelength solution is fixed to that of the KPF data. The relative $\chi^2$ and parameter convergence criteria are set to $10^{-10}$.

\section{Supplementary plots}

\begin{figure*}[t]
\centering
    \includegraphics[width=\linewidth]{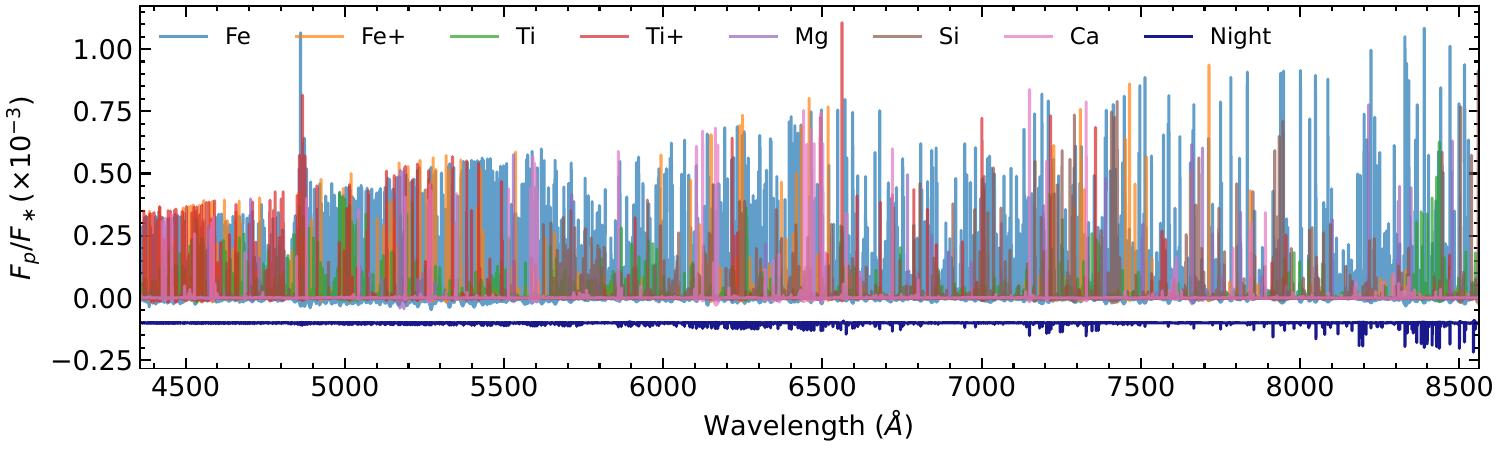}
    \caption{High-pass filtered spectral templates of individual species used in cross-correlation analyses. The models are generated with the best-fit parameters from retrievals (Section~\ref{sec:analysis_retrieval}). The navy line shows a nightside model spectrum with all species for comparison.
    \label{fig:template}}
\end{figure*}

We show the template spectra for various species used for the cross-correlation analysis in Fig.~\ref{fig:template}.  

\begin{figure}[t]
\centering
    \includegraphics[width=0.5\linewidth]{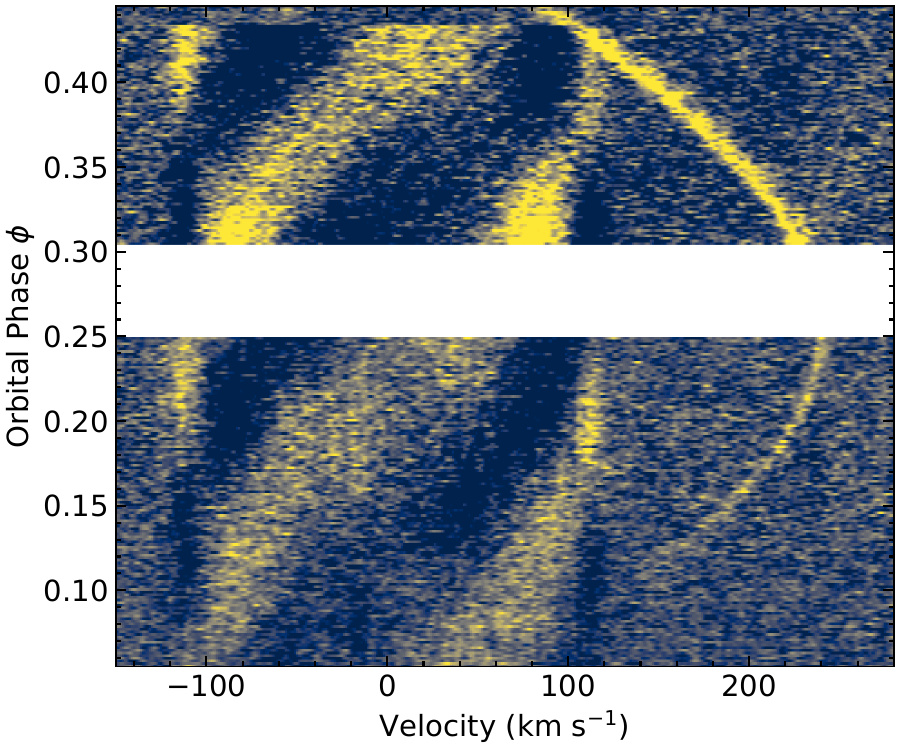}
    \caption{Same as the CCF map in Fig.~\ref{fig:ccf_phase} but before the correction for stellar pulsation pattern.
    \label{fig:ccf_pulsation}}
\end{figure}

We present the CCF map before correcting for the stellar pulsation with Gaussian Processes in Fig.~\ref{fig:ccf_pulsation}. 
The stellar contamination overlays the planetary emission signal for orbital phases $\phi<0.1$ and $\phi>0.4$.

\begin{figure}[t]
\centering
    \includegraphics[width=0.5\linewidth]{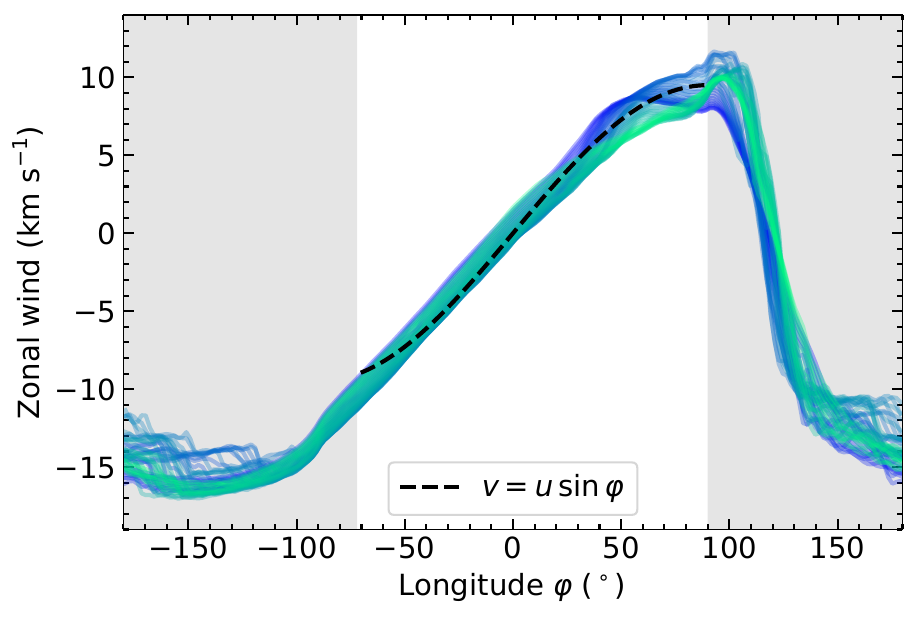}
    \caption{Zonal wind speed across longitudes from the GCM result with $\tau_\mathrm{drag}=10^7$ s. The colored lines represent various latitudes between $\pm 45^\circ$. 
    The gray shaded region represents the nightside longitudes, while the the white region represents the dayside longitudes ($-70^\circ<\varphi<90^\circ$) covered by the observed phases. The black dashed line shows a sinusoidal curve that well represents the wind pattern predictions from GCMs. This also justifies our parametrization of winds in Section~\ref{sec:analysis_retrieval}.
    \label{fig:u_phi}}
\end{figure}

With the GCM presented in Section~\ref{sec:gcm}, we find that the zonal winds across the dayside hemisphere can be well described by a sinusoidal function of longitude as shown in Fig.~\ref{fig:u_phi}.

\begin{figure}[t]
\centering
    \includegraphics[width=\linewidth]{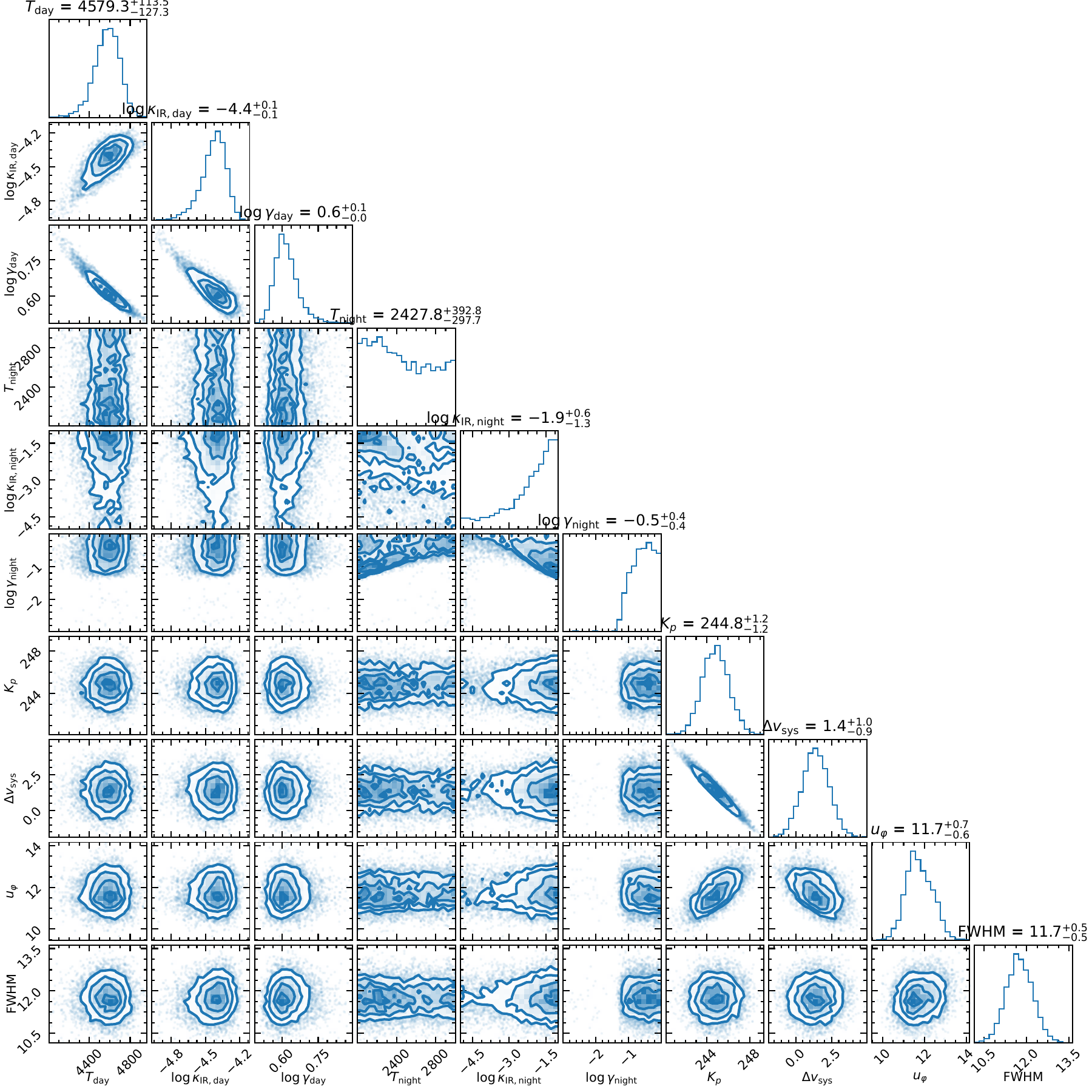}
    \caption{Corner plot of retrieved posterior distributions of parameters as listed in Table~\ref{tab:params}. The median values and 1$\sigma$ uncertainties are summarized on top.
    \label{fig:corner_full}}
\end{figure}

Finally, we show the full corner plot of our retrieval analysis in Fig.~\ref{fig:corner_full}.

\bibliography{manuscript}
\bibliographystyle{aasjournal}

\end{document}